\documentclass[a4paper,12pt]{article}
\pdfoutput=1 
\usepackage{jheppub} 

\usepackage[T1]{fontenc} 
\usepackage{physics,tensor,caption,subcaption}
\usepackage{float}
\def \cH {H}
\def \cR {\mathcal{R}}

\def \Lstar {L_{\star}}
\def \Lamstar {\Lambda_{\star}}

\title{Small free field inflation in higher curvature gravity}

\author[a,b]{Jos\'e D. Edelstein,}
\author[c,d]{Robert B. Mann,}
\author[a,b]{David V\'azquez Rodr\'\i guez,}
\author[a,b]{Alejandro Vilar L\'opez}
\affiliation[a]{Departamento de F\'\i sica de Part\'\i culas, Universidade de Santiago de Compostela, E-15782 Santiago de Compostela, Spain}
\affiliation[b]{Instituto Galego de F\'\i sica de Altas Enerx\'\i as (IGFAE), Universidade de Santiago de Compostela, E-15782 Santiago de Compostela, Spain}
\affiliation[c]{Department of Physics and Astronomy, University of Waterloo, Waterloo, Ontario, N2L 3G1, Canada.}
\affiliation[d]{Perimeter Institute, 31 Caroline St. N., Waterloo, Ontario, N2L 2Y5, Canada.}
\emailAdd{jose.edelstein@usc.es}
\emailAdd{rbmann@uwaterloo.ca}
\emailAdd{davidvazquez.rodriguez@usc.es}
\emailAdd{alejandrovilar.lopez@usc.es}

\abstract{Within General Relativity, a minimally coupled scalar field governed by a quadratic potential is able to produce an accelerated expansion of the universe provided its value and excursion are larger than the Planck scale. This is an archetypical example of the so called large field inflation models. We show that by including higher curvature corrections to the gravitational action in the form of the Geometric Inflation models, it is possible to obtain accelerated expansion with a free scalar field whose values are well below the Planck scale, thereby turning a traditional large field model into a small field one. We provide the conditions the theory has to satisfy in order for this mechanism to operate, and we present two explicit models illustrating it. Finally, we present some open questions raised by this scenario in which inflation takes place completely in a higher curvature dominated regime, such as those concerning the study of perturbations.}  

\begin{document}
\maketitle

\section{Introduction}
\label{sec-Introduction}

The coarse-grained universe is homogeneous and isotropic to a great precision. This led to the hypothesis of cosmological inflation, which has been increasingly supported by experimental data. In the framework of General Relativity, it has been necessary to introduce an agent responsible for inflation. The simplest versions are based on a scalar field, the inflaton, slowly rolling down its potential. There now exist a broad range of possible scenarios with differing degrees of consensus. Sticking to the simplest setups they typically fall into two categories: large field and small field inflation, depending on the comparison of the scalar field, $\phi$, and its excursion in field space, $\Delta\phi$, with the Planck scale \cite{Senatore:2016aui}. The excursion, in particular, is linked to the tensor-to-scalar ratio, $r$, through the Lyth bound \cite{Lyth:1996im}, thereby apparently linking  their fates. Prominent examples of both possibilities are given respectively by the quadratic potential and so-called hilltop inflation \cite{Boubekeur:2005zm}.

In spite of the fact that evidence points towards curvature during inflation being large, the study of higher curvature corrections to General Relativity in this context has been customarily disregarded since the cosmic expansion rate, $H = \dot a/a$, and the physical wave number, $k/a$, at horizon exit are likely to be much less than $M_{\rm Pl} \simeq 2.4 \times 10^{18}$ GeV \cite{Weinberg:2008hq}. However, it is far less clear whether these quantities are negligible when compared with the characteristic scale $\Lstar^{-1}$ of the putative theory underlying inflation, given that it can be well below the Planck scale. For instance, in String Theory corrections to General Relativity are suppressed by the string scale, $\alpha^\prime \sim \Lstar^2$, which can be several orders of magnitude larger. 

We want to be agnostic in the present work and explore possible consequences that may result from the resummation of the whole series of higher curvature corrections to General Relativity,
\begin{equation}
\mathcal{I}_{\rm grav} = \frac{M_{\rm Pl}^2}{2} \int \mathrm{d}^4 x \sqrt{-g} \left[ R + \sum_{n=3}^{\infty} c_n \Lstar^{2n-2} \cR_{(n)} \right] ~,
\label{GravityAction}
\end{equation}
regardless of their origin. Our approach will be bottom-up: we will consider $n$-th order higher curvature densities $\cR_{(n)}$ constructed from contractions of the metric and the Riemann tensor, and complying order by order with the following criteria:
\begin{enumerate}
\item The equations of motion, when linearized around any maximally symmetric spacetime, are second order. For these backgrounds, the only propagating degree of freedom is the usual Einstein graviton \cite{Bueno:2016xff,Hennigar:2016gkm,Bueno:2016ypa}.
\item There is a smooth connection with General Relativity in the limit $c_n \rightarrow 0$. This means, in particular, that we would choose the {\it right} vacuum if there were many (see \cite{Camanho:2011rj} for a similar discussion in the context of Lovelock theory).
\item Solutions to the field equations admit non-hairy deformations of the Schwarzschild black hole with well-behaved thermodynamic properties (and also Taub-NUT/bolt solutions) characterized by a single function, $g_{tt} g_{rr}=-1$, which satisfies a second-order ordinary differential equation \cite{Bueno:2016xff,Hennigar:2017ego,Bueno:2018uoy}. 
\item On  Friedmann-Lema\^itre-Robertson-Walker (FLRW) backgrounds, the equations of motion for the scale factor $a(t)$ are second-order, thus providing sensible cosmological models \cite{Arciniega:2018fxj,Arciniega:2018tnn}.
\end{enumerate} 
The first non-trivial\footnote{The quadratic curvature invariant, $\cR_{(2)}$, is bound to be the Lanczos-Gauss-Bonnet combination \cite{Bueno:2016xff}, which is topological in four dimensions.} density, $\cR_{(3)}$, was identified in \cite{Arciniega:2018fxj}, and it was soon after shown how to construct all of them \cite{Arciniega:2018tnn}. Interestingly enough, the inclusion of radiation ---actually, any barotropic fluid for that matter--- triggers a novel mechanism of cosmological accelerated expansion of the universe called Geometric Inflation, without the need to invoke a scalar field (see, also, \cite{Arciniega:2019oxa}). Several features of this scenario were considered in \cite{Cisterna:2018tgx,Erices:2019mkd,Arciniega:2020pcy,Pookkillath:2020iqq}.

In a recent paper, it was shown that the amount of radiation necessary to account for at least $60$ e-folds of inflation is exceedingly large \cite{Edelstein:2020nhg}. Invoking a scalar field marginally solves this problem, on generic grounds, by producing a cascading scenario in which an epoch of Geometric Inflation terminates smoothly into a last stage that is nothing but the familiar inflaton setup of inflation in General Relativity. Remarkably, the scalar field remains almost frozen during the cascading process, which allows  reduction of its initial value, $\tilde\phi$, and its excursion, $\Delta\phi$. Yet, large field scenarios as free field inflation ---{\it i.e.}, those triggered by a scalar field with a quadratic potential--- remain large field. This means that issues raised by a transplanckian excursion, $\Delta\phi > M_{\rm Pl}$, would have to be dealt with, such as the unknown details of the scalar field potential coming from integrating out very massive fields to which the inflaton may couple in the UV complete theory. 

In this paper we will provide an unexpected way out of this dilemma. We will show that the archetypical example of large field inflation, which is the scalar field with a quadratic potential,
\begin{equation}
\mathcal{I}_{\rm scalar} = \int \mathrm{d}^4 x \sqrt{-g} \left[  - \frac{1}{2} \left( \nabla \phi \right)^2 - \frac{1}{2} m^2 \phi^2 \right] ~ .
\label{ScalarFieldAction}
\end{equation}
can be converted into a small field scenario when coupled to \eqref{GravityAction}.\footnote{Albeit apparently unrelated to our approach, it is interesting to notice that recent papers show that a similar conversion can be achieved by dealing with the Starobinsky model in the framework of the Palatini formalism \cite{Enckell:2018hmo,Antoniadis:2018ywb}.} Contrary to the models discussed in \cite{Edelstein:2020nhg}, here we will not need to include radiation; the action is entirely given by $\mathcal{I}_{\rm grav} + \mathcal{I}_{\rm scalar}$. We will work in a flat Friedmann-Lema\^itre-Robertson-Walker (FLRW) spacetime\footnote{Spatial curvature can be easily included, as mentioned in \cite{Arciniega:2018tnn}. Similarly, a cosmological constant can be incorporated in \eqref{GravityAction}, but it should be irrelevant for early time inflation.}
\begin{equation}
ds^2 = -dt^2 + a(t)^2 \left( dr^2 + r^2 d\Omega^2 \right) ~,
\label{FLRW}
\end{equation}
where the associated generalized Friedmann equations for the scale factor $a(t)$,
\begin{eqnarray}
3 F(H) &=& \frac{1}{M_{\rm Pl}^2} \rho ~, \label{FriedmannEq} \\ [0.5em]
- H^\prime\,\frac{dF(H)}{dH} &=& \frac{1}{M_{\rm Pl}^2} (\rho + P) ~,
\label{FriedmannEq2}
\end{eqnarray}
are second-order\footnote{In order to study cosmological inflation, it is more suitable to trade the time coordinate variable for the number of e-folds, $N$. In terms of the latter, the scale factor behaves as $a = \tilde{a}\,e^N$, where $\tilde{a}$ is the initial value; thereby
\begin{equation}
\frac{\mathrm{d}}{\mathrm{d}t} = H \frac{\mathrm{d}}{\mathrm{d}N} ~.
\label{dNefolds}
\end{equation}
We use primes to denote derivatives with respect to $N$ and dots for the usual derivatives with respect to the cosmological time. We implicitly assume $\dot{a} > 0$, which will certainly be the case for inflationary models.} and entirely given in terms of a single function
\begin{equation}
F(H) \equiv H^2 + \Lstar^{-2} \sum_{n=3}^{\infty} (-1)^n c_n \left( \Lstar H \right)^{2n} ~,
\label{F}
\end{equation}
where $H\equiv \dot{a}/a$ is the usual Hubble parameter. Here, $\rho$ and $P$ are the density and pressure obtained from the kinetic and potential contributions to the matter energy-momentum tensor, $\rho = K_\phi + V_\phi$, and $P = K_\phi - V_\phi$, where
\begin{equation}
K_\phi = \frac{1}{2} H^2 \phi'^2 ~ , \qquad V_\phi = \frac{1}{2} m^2 \phi^2 ~,
\label{KphiVphi}
\end{equation}
and the reduced Planck mass is $M_{\rm Pl}^2 = (8\pi G)^{-1}$. Notice that, in this scheme, the conservation equation for the matter energy-momentum tensor reads 
\begin{equation}
\left( H \phi' \right)' + 3 H \phi' + \frac{m^2}{H} \phi = 0  ~,
\label{KGEquation}
\end{equation}
which is consistent with \eqref{FriedmannEq} and \eqref{FriedmannEq2}. As it is customary in inflationary cosmology, we will work in units $\hbar = c = 1$, explicitly writing factors of the (reduced) Planck mass, $M_{\rm Pl}$.

The coefficients $c_n$ in \eqref{GravityAction} are expected to be computed from a UV complete theory of gravity. Indeed, it was recently observed that T-duality is stringent enough to allow for a complete classification of duality invariant $\alpha^\prime$ corrections in a cosmological setup \cite{Hohm:2019jgu}. The Friedmann equations (in the string frame) can be expressed in terms of a single function that looks exactly like \eqref{F}, except for the fact that it does not exclude a quartic term. This was used to show that there are $T$-duality invariant theories featuring string frame\footnote{In the Einstein frame, $F(H)$ must satisfy a second order non-linear ordinary differential equation whose solutions lead to cosmologies with a constant dilaton and power law scale factors \cite{Krishnan:2019mkv}.} de Sitter vacua that are non-perturbative in $\alpha^\prime$ provided $F(H)$ satisfies some simple properties \cite{Hohm:2019ccp}. String Theory is certainly invariant under $T$-duality,\footnote{The space of $T$-duality invariant theories seems broader, though \cite{Edelstein:2019wzg}.} thus providing a concrete and rigorous scenario where the coefficients $c_n$ may be computed, at least in principle. It is conceivable that there may be other constructions leading to the determination of these coefficients or to a differential equation for $F(H)$ (as in \cite{Krishnan:2019mkv}).

Leaving aside the origin of $F(H)$, whether it originates in a bottom-up construction or comes from a putative UV complete quantum theory of gravity, we will consider $\mathcal{I}_{\rm grav} + \mathcal{I}_{\rm scalar}$, and show that a novel scenario emerges if it satisfies simple and quite generic properties. Most importantly, we will introduce a new parameter, $\Lambda_{\rm inf}$, which establishes the energy scale of inflation, and the regime of interest will emerge when $\Lambda_{\rm inf} \gg \Lstar^{-1}$. We emphasize that $\Lstar$ can be constrained by astrophysical tests. For instance, Shapiro time delay experiments in the solar system lead to $\Lstar \lesssim  10^{8}$m, if performed in the cubic or quartic theories \cite{Hennigar:2018hza,Khodabakhshi:2020hny}.  We will demonstrate that under these circumstances cosmological inflation might have taken place {\it far away} from the General Relativity regime. This would represent a major departure from the standard inflationary setup. In particular, we show that a paradigmatic large field inflationary model as provided by \eqref{ScalarFieldAction} becomes a small field model; both the initial value of the scalar field and its excursion drop to subplanckian values, $\tilde\phi$ and $\Delta\phi \ll   M_{\rm Pl}$. These models are potentially able to provide a satisfactory inflationary evolution, at least at the background level.\footnote{The study of perturbations is quite involved in these theories but definitely needs to be addressed in order to consider them viable.}

The paper is organized as follows. In section 2 we study the properties that $F(H)$ must satisfy in order to favor small field inflation. We will show that a novel scenario arises if $\Lambda_{\rm inf} \gg \Lstar^{-1}$ and $F(H)$ is a steep function around $H \sim \Lambda_{\rm inf}$. Section 3 is devoted to the study of two models fulfilling these conditions. We show that the archetypical example of large field (chaotic) inflation, {\it i.e.}, the scalar field with a quadratic potential, produces a quasi-de Sitter expansion with a large number of e-folds for subplanckian values of the scalar field and its excursion. Section 4 contains further discussions, future prospects, and conclusions.

\section{Properties of $F(H)$ to produce inflation}
\label{sec-Properties}

As mentioned in the introduction, it was recently pointed out  \cite{Arciniega:2018fxj,Arciniega:2018tnn} that theories of the form \eqref{GravityAction} are able to produce accelerated expansion on a quite general basis. Let us pause for a moment to understand why this is so in the presence of a generic form of matter, since this analysis will provide us with the tools needed to develop useful scalar field models. The $\epsilon$ parameter is the conventional measure of accelerated expansion, satisfying $\epsilon < 1$ whenever $\ddot{a} > 0$ and $\epsilon > 1$ if $\ddot{a} < 0$. Using the generalized Friedmann equations \eqref{FriedmannEq} and \eqref{FriedmannEq2} we obtain:
\begin{equation}
\epsilon = - \frac{H'}{H} = \frac{3 F(H)}{H\,\mathrm{d}F/\mathrm{d}H} \left( 1 + \frac{P}{ \rho} \right) ~.
\label{GeneralEpsilon}
\end{equation}
Suppose now the matter satisfies a barotropic equation of state of the form $P = w \rho$, and the function $F$ behaves roughly as some power $H^{2k}$. In this case:
\begin{equation}
\epsilon = \frac{3}{2k} \left(1 + w \right) ~.
\label{EpsilonOriginalGeometricInflation}
\end{equation}
This was the situation in the original cubic model of geometric inflation \cite{Arciniega:2018fxj}, where, at early times, the higher-order term dominates due to the large value of $H$ and $F(H) \sim H^6$. In that case, we get $\epsilon < 1$ even for radiation, with $w = 1/3$. Similar considerations can be applied to the more general models of \cite{Arciniega:2018tnn}, where terms with larger $k$ are included, which only help diminishing $\epsilon$.

All this can be summarized in a single statement by looking at \eqref{GeneralEpsilon}: a fast-growing $F$ will help to produce accelerated expansion. Let us make this more precise for the case of a scalar field. Using the definition of $K_{\phi}$ and $V_{\phi}$ as the kinetic and potential energies of the scalar \eqref{KphiVphi}, so that $\rho + P = 2 K_{\phi}$, the $\epsilon$-parameter is:
\begin{equation}
\epsilon = \frac{6 F(H)}{H \mathrm{d}F/\mathrm{d}H} \frac{K_{\phi}}{K_{\phi} + V_{\phi}} ~.
\label{EpsilonScalar}
\end{equation}
Now we can ask the following question: in terms of the scalar field energies, when does inflation end in these models? This is easily answered by imposing the condition $\epsilon < 1$ to have accelerated expansion,
\begin{equation}
\frac{V_{\phi}}{K_{\phi}} > \frac{6 F(H)}{H \mathrm{d}F/\mathrm{d}H} - 1  ~.
\label{EndOfScalarInflationEnergies}
\end{equation}
This gives the well-known result $V_{\phi} > 2 K_{\phi}$ to have accelerated expansion in General Relativity. When higher-order terms are included, instead, a fast growing $F$ decreases the limiting ratio, and eventually it can become irrelevant for any model with a positive definite potential.

Alternatively, suppose $V_{\phi} \leq \alpha K_{\phi}$, and we require inflation to take place over some range of $H\in (H_1, H_2)$. This implies
\begin{equation}
\frac{6}{1 + \alpha} \frac{F(H)}{H \mathrm{d}F/\mathrm{d}H} \leq \epsilon < 1 \qquad \Rightarrow \qquad F(H) > F(H_1) \left( \frac{H}{H_1} \right)^{6/(1+\alpha)}
\end{equation}
as a condition on $F(H)$. Conversely, this implies that provided
\begin{equation*}
F(H) \leq F(H_1) \left( \frac{H}{H_1} \right)^{6/(1+\alpha)} \Rightarrow \quad \text{it is impossible to inflate with } V_{\phi} \leq \alpha K_{\phi}  ~.
\end{equation*}
As an example, in General Relativity the inequality is true with $\alpha = 2$. Therefore, it is impossible to inflate with $V_{\phi} \leq 2 K_{\phi}$.

\subsection{Connection with the GR regime}
\label{sec-Connection}

All the previous arguments point towards the fact that a fast-growing $F$ benefits inflation. But we are constrained by the fact that $F(H) \sim H^2$ when $H \to 0$ in order to obtain a sensible GR limit to \eqref{GravityAction}. This GR behavior of the function $F$ is at the root of the large field values of the scalar field needed to inflate in conventional models. Let us quickly review how this works. 

We will assume throughout this section that we are in a slow-roll regime, which means we neglect the kinetic contribution to the energy density and also its derivative,\footnote{Explicitly, the conditions to obtain the slow-roll equations are $K_{\phi} \ll V_{\phi}$, and $|K'_{\phi}| \ll |V'_{\phi}|$. In terms of the field, $H^2 \phi'^2 \ll m^2 \phi^2$ and $|(H \phi')'| \ll m^2 |\phi| / H$.} leaving the following evolution equations:
\begin{equation}
F(H) \simeq \frac{m^2}{6 M_{\rm Pl}^2} \phi^2 ~ , \qquad 3 H^2 \phi' \simeq - m^2 \phi ~ .
\label{SlowRollEquations}
\end{equation}
This allows us to obtain analytic results that are reasonably accurate and can also be verified numerically. Within this regime,
\begin{equation}
\epsilon \simeq \frac{2 m^2}{3 H^2} \frac{F(H)}{H \mathrm{d}F/\mathrm{d}H} ~ .
\label{EpsilonSlowRoll}
\end{equation}
In GR, $F(H) = H^2$, thereby $6 M_{\rm Pl}^2 H^2 \simeq m^2 \phi^2$, which leads to the familiar expression $\epsilon \simeq 2 M_{\rm Pl}^2 / \phi^2$; therefore $\phi > \sqrt{2} M_{\rm Pl}$ to have inflation.\footnote{One could wonder whether the slow-roll approximation is an essential feature of this bound. It turns out it is not: since in GR one must always have $V_{\phi} > 2 K_{\phi}$ to produce inflation, potential energy dominates in any accelerated expansion phase. More accurate numerical results can be obtained in different situations, and in the quadratic potential model one always has to run through transplanckian field values to have a significant number of e-folds of inflation.}

We have seen that a fast-growing $F(H)$ could potentially guarantee inflation irrespective of the field evolution. This follows from \eqref{EndOfScalarInflationEnergies}: if $F(H)$ grows (at least) as $H^6$, then the condition for accelerated expansion becomes   trivial, $V_{\phi} \geq 0$. Motivated by this, let us study, in the slow-roll regime, a simple model of the following form:
\begin{equation}
F(H) = H^2 + \beta_k H^{2k} ~ ,
\label{GRAndPowerLawModel}
\end{equation}
with $k \geq 3$. Let us also assume that we start in a regime in which the GR term is irrelevant, so that we can approximate $F(H) \approx \beta_k H^{2k}$. This would be the situation in which we expect inflation to be guaranteed. The slow-roll equations \eqref{SlowRollEquations} can be analytically solved, giving:
\begin{equation}
H \approx m \gamma_k^{-1/(2k)} \left( \frac{\phi}{M_{\rm Pl}} \right)^{1/k} ~, \qquad \frac{\phi}{M_{\rm Pl}} \approx \left[ \left( \frac{\tilde{\phi}}{M_{\rm Pl}} \right)^{2/k} - \frac{2}{3k} \gamma_k^{1/k} N \right]^{k/2} ~,
\label{PowerLawSlowRollSolution}
\end{equation}
where $\tilde{\phi}$ is the initial value of the field at $N=0$, and we have defined the dimensionless combination $\gamma_k \equiv 6 \beta_k m^{2(k-1)}$. There are two ways in which this slow-roll evolution can stop. 

One is if we break the slow-roll conditions, because then the field rapidly falls to the minimum of the potential at $\phi = 0$, dissipating energy and forcing connection with GR. We can estimate the value of the field for which this happens by looking, with the result \eqref{PowerLawSlowRollSolution}, at the conditions $K_{\phi} \approx V_{\phi}$ and $|K'_{\phi}| \approx |V'_{\phi}|$. It turns out that the one producing a larger field value is the first one, which happens at $\phi \approx \phi_1$,
\begin{equation}
\phi_1 = \frac{M_{\rm Pl}}{9^{k/2}} \sqrt{\gamma_k} ~ .
\label{Condition1StopInflation}
\end{equation}
Notice that, for small but non-zero $\gamma_k$, this would mean that the slow-roll approximation is good up until very small field values. In addition, it is not difficult to show that the number of e-folds needed to reach this value of the field is
\begin{equation}
N_1 \approx \frac{3 k}{2 \gamma_k^{1/k}} \left( \frac{\tilde{\phi}}{M_{\rm Pl}} \right)^{2/k} - \frac{k}{6} ~ .
\end{equation}
This can be made arbitrarily large for sufficiently small $\gamma_k$, irrespective of the initial $\tilde{\phi}$. But if we have a function $F(H)$ such as \eqref{GRAndPowerLawModel}, in which the GR term is present, the solution \eqref{PowerLawSlowRollSolution} can also cease  to be valid if, even within the slow-roll regime, we abandon the higher-order domination phase, so that the approximation $F(H) \approx \beta_k H^{2k}$ no longer holds. This happens roughly when $\beta_k H^{2(k-1)} \approx 1$, which, in terms of the scalar field, means $\phi \approx \phi_2$,
\begin{equation}
\phi_2 = \frac{6^{k/[2(k-1)]}}{\gamma_k^{1/[2(k-1)]}} M_{\rm Pl} ~ .
\label{Condition2StopInflation}
\end{equation}
Contrary to $\phi_1$, this increases when $\gamma_k$ decreases. So, for small $\gamma_k$, GR would take control of the evolution within the slow-roll regime, but we know in that case the field cannot decrease below $\sqrt{2} M_{\rm Pl}$ while still inflating. In fact, it is not hard to show that the value of $\gamma_k$ which makes $\phi_1 = \phi_2 = \sqrt{2/3} \, M_{\rm Pl}$ is $\gamma_k = 6 \cdot 9^{k-1}$. This is in a sense   optimized: we cannot decrease the field below this value whilst keeping inflationary evolution. For smaller $\gamma_k$, we enter the GR regime before reaching this optimum, which means inflation has to stop at $\phi = \sqrt{2} M_{\rm Pl}$. For larger $\gamma_k$, we abandon the slow-roll regime within higher-order domination, but then dissipation quickly reduces the energy density and we return to GR domination, also spoiling inflation.

This example, based on the particular function \eqref{GRAndPowerLawModel}, was intended to demonstrate that the connection of $F(H)$ with the GR regime makes it very difficult to  significantly decrease the values of the scalar field needed to produce inflation. But the fact that we were not able to do it also points towards an alternative that  may be successful. All we need to have a large number of e-folds of inflation with a small field is a region in which the function $F(H)$ grows fast, like the situation leading to the slow-roll solution \eqref{PowerLawSlowRollSolution}. Then, to avoid the problem arising when connecting to the GR part, $F(H) \sim H^2$, we can try to introduce an intermediate regime, separating the region of fast-growing $F(H)$ from the one in which it approaches the GR quadratic expression. 

In the next section we shall construct two such particular models, leading to numerical results that validate our expectation.

\section{Explicit models}
\label{sec-Models}

Motivated by the previous discussion, we present here two explicit models that implement the idea we have just described. The models differ enough so that they highlight the key similar features (as opposed to the particular form of $F(H)$) that illustrate what is required to obtain inflation with scalar field values well below the Planck scale. Recall the basic properties we require of $F(H)$ are as follows.
\begin{enumerate}
\item For small $H$, $F(H) \sim H^2 + \mathcal{O}(H^6)$. This is forced by the general expression \eqref{F}, and the fact that we cannot have an $H^4$ term in four dimensions for GQTG gravity \cite{Hennigar:2017ego}.
\item We require an almost flat region of $F(H)$, whose only purpose is to separate the inflating part of the function from the GR one. This circumvents the problems mentioned in the previous section, in which direct connection with GR forbade inflation with small field values.
\item Finally, we require a region in which $F(H)$ grows reasonably fast, and where inflation will be produced. The mechanism will be similar to the one discussed after \eqref{PowerLawSlowRollSolution} for the simple, pure higher-order model.
\end{enumerate}
Our first example will be provided by the following \emph{power-law model}:
\begin{equation}
F(H) = H^2  \left( \frac{1}{1 +  H^2/\Lamstar^2} + \frac{H^2/\Lamstar^2}{1 + H^4 / \Lamstar^4} \right) \left( 1 + \frac{H^4}{\Lambda_{\rm inf}^4} \right) ~.
\label{PowerLawModel}
\end{equation}
Despite its seemingly arbitrary appearance, this model has two free parameters: $\Lamstar = L_\star^{-1}$ and $\Lambda_{\rm inf}$, and implements in a simple way the aforementioned conditions. The idea is that $\Lambda_{\rm inf} \gg \Lamstar$, so that we find the three regimes previously mentioned. For $H < \Lamstar$, we get the GR regime, with $F(H) \sim H^2 + \mathcal{O}(H^6)$. For $\Lamstar < H < \Lambda_{\rm inf}$, we enter the flat part of the function. Finally, approaching $\Lambda_{\rm inf}$ we start to enter the inflating phase in which the function grows as a quartic power. Essentially, $\Lamstar$ marks the energy scale of corrections to GR, while $\Lambda_{\rm inf}$ marks the energy scale of inflation.

An alternative expression for $F(H)$ is given by the following \emph{Gaussian model}:
\begin{equation}
F(H) = \left[ \frac{H^2}{2 + H^2/\Lamstar^2} + \left( 1 - \frac{1}{\Gamma c_{\sigma}^2} \right) \frac{\cH^4/(4\Lamstar^2)}{1 + \cH^4/\Lamstar^4} \right] \left[ 1 + e^{\frac{1}{2 c_{\sigma}^2}} e^{- \frac{(\cH^2 / \Lamstar^2 - \Gamma)^2}{2 \Gamma^2 c_{\sigma}^2}} \right] ~ .
\label{GaussianModel}
\end{equation}
In this case we have three parameters. The first one, $\Lamstar$, plays the same role as before, setting the energy scale at which corrections to GR appear. Then we have two extra parameters, $\Gamma$ and $c_{\sigma}$,  that control the Gaussian factor.  We will assume $\Gamma \gg 1$ and $c_{\sigma} \lesssim 1$. This ensures that inflation here will take place in the growing part of the Gaussian. Its center (in $H$) is located at $\Lambda_{\rm inf} = \sqrt{\Gamma} \Lamstar$, so that $\Gamma \gg 1$ guarantees we are separating the inflationary regime from the GR one, which stops, as mentioned, around $\Lamstar$. Similarly, $c_{\sigma}$ determines the relative width of the Gaussian\footnote{The center of the Gaussian in $H^2$ is at $\Lambda_{\rm inf}^2 = \Gamma \Lamstar^2$. The standard deviation is $\sigma_{H^2} = c_{\sigma} \Gamma \Lamstar^2$. Thus, $c_{\sigma} = \sigma_{H^2} / \Lambda_{\rm inf}^2$.}, thus we set $c_{\sigma} \lesssim 1$. Notice that this function has a maximum value at the center of the Gaussian, assuming $\Gamma \gg 1$. There is then a maximum allowed value of the energy density; otherwise there is no solution to the first Friedmann equation  \eqref{FriedmannEq}. This also means that, for certain values of $\rho$, there are two solutions for $H$. We will not care about this issue here, choosing always to work with the first solution which connects with the GR regime. 

\subsection{Power-law model: numerical results}
\label{sec-PowerLawModel}

\begin{figure}[ht]
\centering
\includegraphics[width=.67\linewidth]{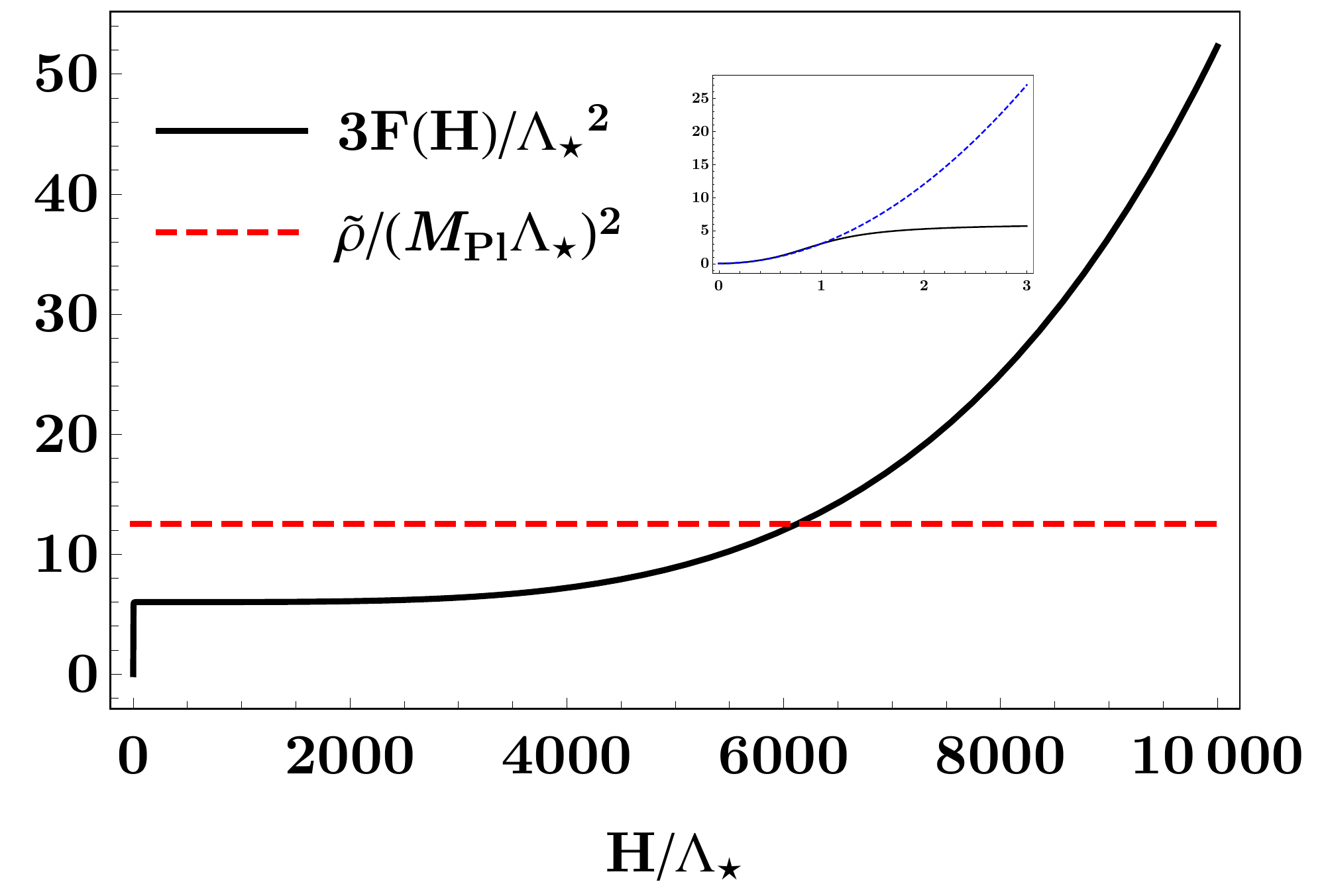}    
\caption{Initial set up in the power law model with parameters $\Lambda_{\rm inf} = 6000 \, \Lamstar$, $m = 500 \, \Lamstar$ and an initial value $\tilde{\phi} = 0.01 \, M_{\rm Pl}$. Inset: Zooming in showing that $F(H)$ approaches the GR value ---blue dashed line---, $H^2$, as $H \to 0$.} 
\label{fig-setuppowerlaw}
\end{figure}

Let us now present an explicit example for the \emph{power-law model}, \eqref{PowerLawModel}. Notice that, from the form of $F(H)$ and the Friedmann equations  \eqref{FriedmannEq} and \eqref{FriedmannEq2}, $H$ and $m$ are naturally measured in units of $\Lamstar$, while the field $\phi$ will be measured in units of $M_{\rm Pl}$. In particular, in figure \ref{fig-setuppowerlaw} we plot the function $F(H)$ for $\Lambda_{\rm inf} = 6000 \, \Lamstar$, and the initial energy density of the scalar field for $m = 500 \, \Lamstar$, $\tilde{\phi} = 0.01\, M_{\rm Pl}$.
Evolution starts precisely at the intersection between the two lines (whose intersection point defines the initial value of the Hubble parameter, $\tilde{H}$), and proceeds rolling down the curve of $F(H)$, decreasing $H$ and the field $\phi$. When $H \sim \Lamstar$, $F(H)$ starts to behave like in GR, which accounts for the sudden drop at small values of $H / \Lamstar$ at the far left of figure~\ref{fig-setuppowerlaw} (we zoom in the $H \lesssim \Lamstar$ region to display such behavior).

Notice that in this discussion $\Lamstar$ is still a free parameter. To keep things below the Planck scale we can for example set $\Lamstar = 10^{-5} \, M_{\rm Pl}$, which safely produces $m = 0.005 \, M_{\rm Pl}$ and $\tilde{H} \sim 0.06 \, M_{\rm Pl}$. In any case, $\Lamstar$ is a free parameter that must be empirically determined by measuring the energy scale at which corrections to GR become relevant. There is a lot of space between current observational constraints, $\Lamstar \gtrsim 10^{-43}\, M_{\rm Pl}$, and the Planck scale, so any value within that window in which GR would receive corrections whilst the physics remains in a semi-classical regime is in principle acceptable.  

\begin{figure}[ht]
\centering
\includegraphics[width=.76\linewidth]{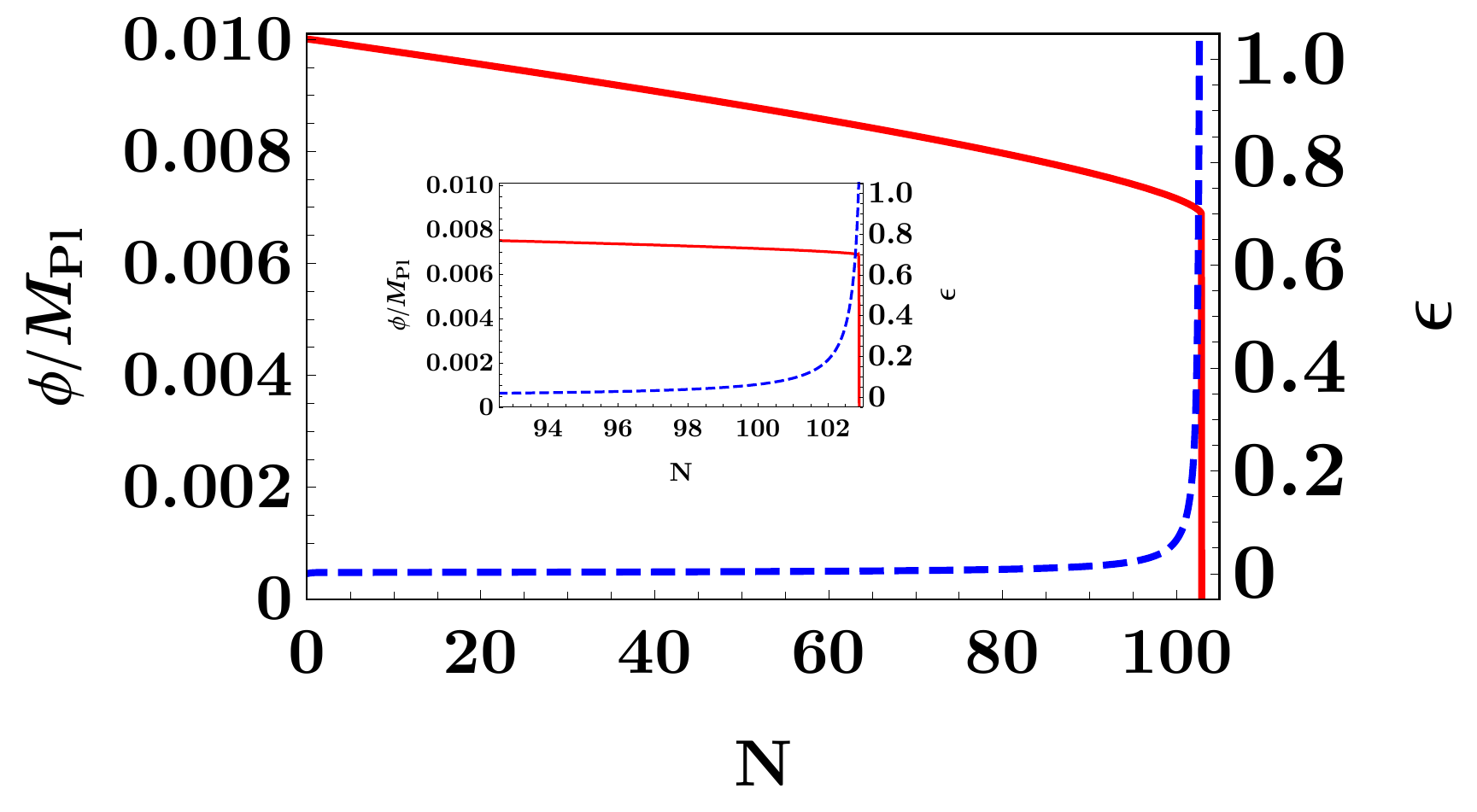}   
\caption{ Scalar field inflationary evolution with initial value $\tilde{\phi}=0.01 M_{\rm Pl}$ and mass $m=500 \, \Lamstar$, producing $N = 103$ e-folds of inflation. Inset: Zooming in over the end of inflation. Notice that the slow-roll condition is valid until the very end, {\it i.e.}, $\epsilon$ is approximately zero during the whole evolution.}
\label{fig-powerlaw}
\end{figure}

After numerically solving the first Friedmann equation \eqref{FriedmannEq} and the scalar field equation \eqref{KGEquation}, we obtain the result plotted in figure \ref{fig-powerlaw}. 
Notice how the field is in a slow-roll regime ---in which $\epsilon$ is negligible--- during a sufficiently large number of e-folds.\footnote{The values here were chosen in order to prove that inflation happens for a significant number of e-folds in a subplanckian scale. It is easy to vary  them; in particular one can get even more e-folds of inflation by increasing the initial value of the field.  For example, $\tilde{\phi} = 0.02 \, M_{\rm Pl}$ with the same profile function and mass produces $N = 645$ e-folds.} After that, the system enters the flat part of $F(H)$, the field decreases rapidly and dissipates a lot of energy, stopping the accelerated expansion and quickly connecting with the GR regime. This realizes the situation discussed at the beginning of this section, in which inflation happens   below the Planck scale whilst completely disconnected from the GR regime.

\subsection{Gaussian model: numerical results}
\label{sec-GaussianModel}

\begin{figure}[ht]
\centering
\includegraphics[width=.67\linewidth]{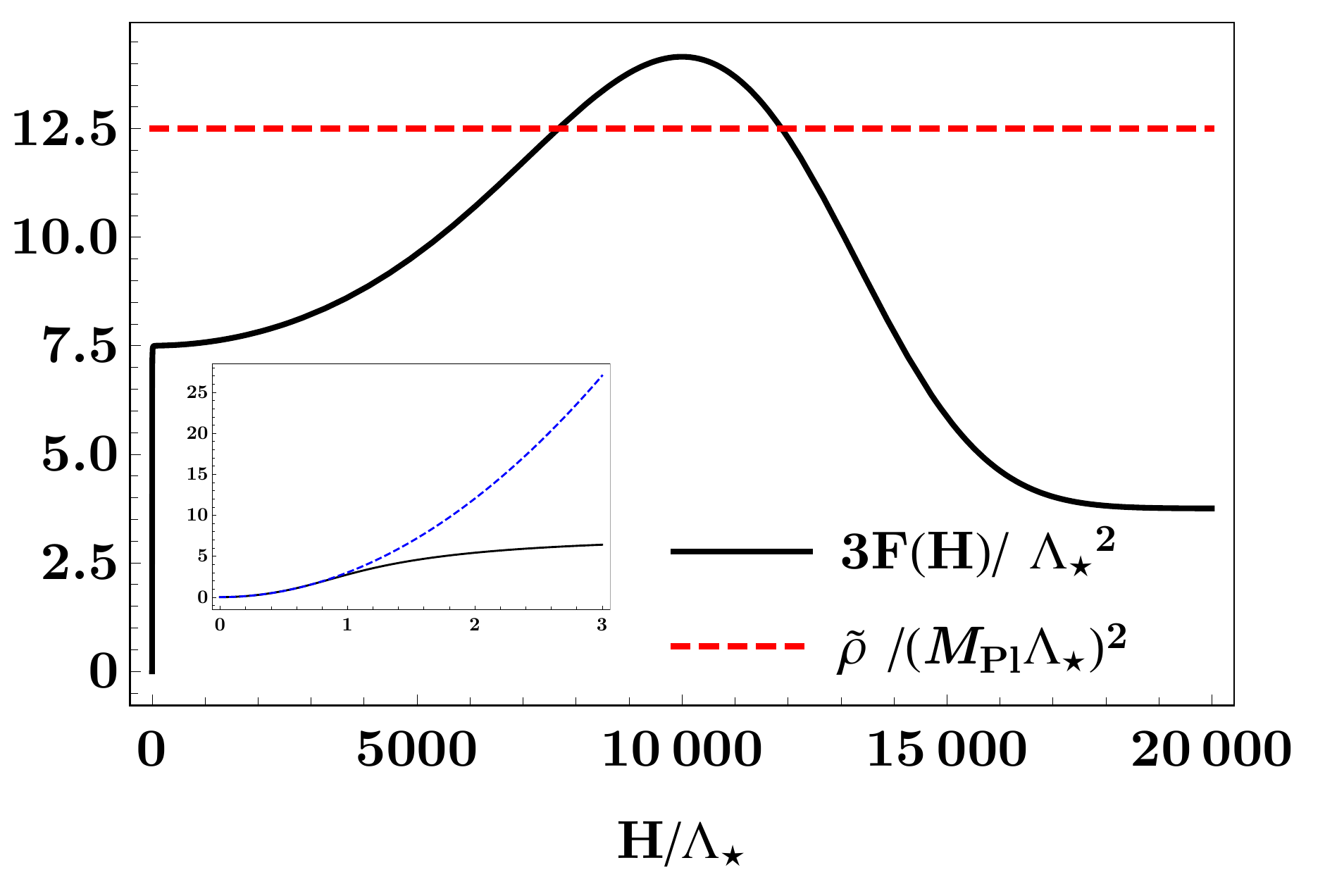}    
\caption{Initial set up in the Gaussian model with parameters $c_\sigma = 0.7$, $\Gamma = 10^8$, $m = 500 \, \Lamstar$, and an initial value $\tilde{\phi} = 0.01 \, M_{\rm Pl}$. Inset: Zooming in showing that $F(H)$ approaches the GR value ---blue dashed line---, $H^2$, as $H \to 0$.} 
\label{fig-setupgaussian}
\end{figure}

An alternative example is provided by the \emph{Gaussian model} \eqref{GaussianModel}, plotted in figure \ref{fig-setupgaussian}.
As mentioned when introducing this model, the maximum value of $F(H)$ makes manifest a degeneracy in the system: for the given initial conditions displayed in the figure, there is a second solution in which $H$ starts to the right of the maximum. We do not consider this branch here, because $\mathrm{d}F / \mathrm{d} H < 0$ produces $H' > 0$, so the Hubble parameter keeps increasing and there is no connection with the General Relativity regime.

\begin{figure}[ht]
\centering
\includegraphics[width=.76\linewidth]{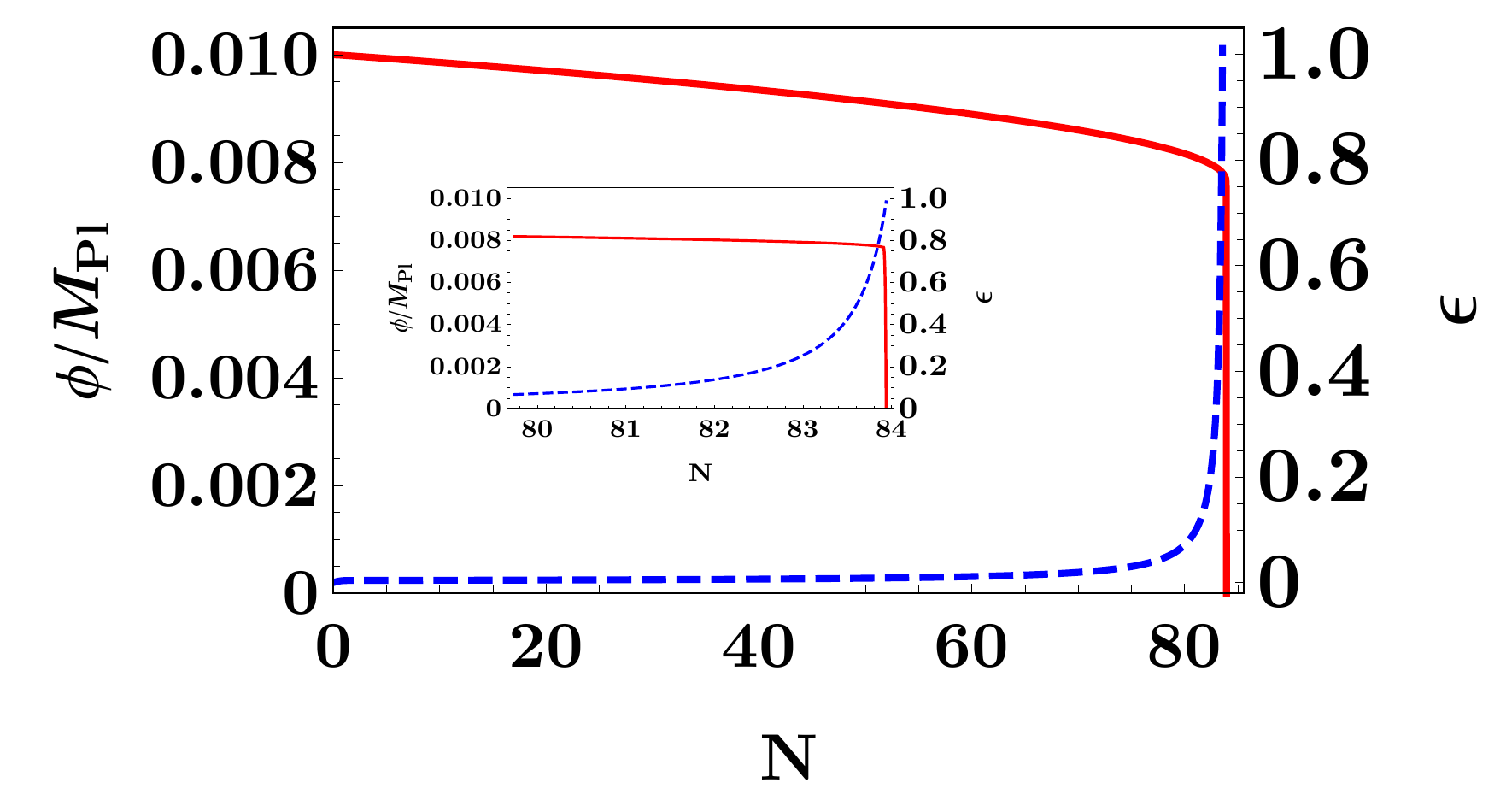}   
\caption{ Scalar field inflationary evolution with initial value $\tilde{\phi}=0.01 M_{\rm Pl}$ and mass $m=500 \, \Lamstar$, producing $N = 84$ e-folds of inflation. Inset: Zooming in over the end of inflation.}
\label{fig-gaussian}
\end{figure}

We start with the same parameters for the field: $\tilde{\phi}=0.01 \, M_{\rm Pl}$, and $m = 500 \, \Lamstar$ than in the previous model, for the sake of comparison. For the function $F(H)$, we set a narrow Gaussian width, $c_\sigma = 0.7$, and a huge factor, $\Gamma = 10^8$, ensuring a large scale separation, $\Lambda_{\rm inf} \gg \Lamstar$. As in the previous model, $\Lamstar$ is a free parameter characterizing the scale of corrections to GR. This time we obtain $N = 84$ e-folds of inflationary evolution before the system enters the flat part of $F(H)$, where once again we get quick dissipation and eventually connection with the GR regime. The numerical results are plotted in figure \ref{fig-gaussian}.
All of the accelerating expansion happens in the fast-growing part of $F(H)$, completely disconnected from the GR regime, and following an almost de Sitter expansion ($\epsilon \approx 0$). Both models display similar features, and it is not hard to convince oneself that this will be the case for any $F(H)$ fulfilling the three properties ---actually, only the last two are strictly relevant--- listed at the beginning of this section.

\section{Discussion and concluding remarks}
\label{sec-Discussion}

We have shown that geometric inflation provides a way out of the transplanckian dilemma of large field inflation. Provided the function $F(H)$ is of the form such that a flat region connects a GR regime with one of rapid growth, it is possible to obtain the requisite number of e-folds to satisfy inflationary phenomenology whilst keeping values of the scalar field and its excursion below the Planck scale. It is possible for $F(H)$ to satisfy these requirements with dependence on as few as two parameters, and no contrivances of the scalar field potential are required.  

Our results suggest that higher-curvature corrections can circumvent transplanckian problems within the semi-classical regime, turning a quintessential model of large field inflation such as the quadratic potential into a small field one, where both the initial value of the field and its range remain below the Planck scale. Let us emphasize that inflation happens completely in the regime in which these corrections are dominant, far from where spacetime is governed by GR. This opens up the possibility to revise all the traditional results of inflation supported by an analysis in which GR is the theory governing the gravitational dynamics, in particular the relation of the tensor-to-scalar ratio $r$ with the excursion of the field given by the Lyth bound \cite{Lyth:1996im} would have to be reconsidered.

This fact also points towards one of the main avenues we would like to explore in the future. Being able to explain the strength of the fluctuations observed in the CMB is a basic requirement for any viable inflationary model. In our theory, in which all higher curvature couplings are turned on, that would mean computing perturbations for the full action \eqref{GravityAction}, an extremely demanding task. Nevertheless, some progress has been made for situations in which only the cubic coupling is present \cite{Cisterna:2018tgx} (see also \cite{EPVLunpublished}). Interestingly enough, from the point of view of the time derivatives, they satisfy second order differential equations. It would be interesting to see whether the particular form of the curvature invariants involved in these theories makes it feasible to compute the spectrum of perturbations for models containing all couplings. We expect the results to be considerably different from the ones obtained in conventional models with the Einstein-Hilbert action ruling the gravitational dynamics, since as we already mentioned inflation happens here in a regime where higher curvature terms are dominant. 

An important problem to address is to study the possible existence of strong instabilities as those found in the cubic theory in the absence of a scalar field \cite{Pookkillath:2020iqq}. Pookkillath, De Felice and Starobinsky studied odd-parity perturbation modes on a spatially homogeneous plane-symmetric Bianchi type I solution of the vacuum equations in the cubic theory \cite{Arciniega:2018fxj}, and showed the presence of (at least) one ghost which triggers a short-time-scale (compared to the Hubble time) classical instability when a small anisotropy develops. The inclusion of a scalar field (and the whole series of higher curvature terms) drastically departs from their analysis and the whole issue must be revisited.

There is a lot more to explore along the road opened by the geometric inflation proposal. The aforementioned study of cosmological perturbations will shed some light onto the degeneracies of the higher-curvature lagrangian densities $\cR_{(n)}$. The possible existence of a network of relations among them via dimensional reduction, as first noticed in \cite{Cisterna:2018tgx} for the cubic case, is worth investigating.

Proposing a semi-classical regime in which gravity has higher-curvature corrections immediately raises some questions, both theoretical and phenomenological. Among the latter, it seems clear that further constraining $\Lamstar$ is necessary. In that respect, the identification of strong gravity phenomena that could serve as labs would make a difference. For instance, the recently suggested possibility that primordial black holes with Earth-like masses may have been captured by the Solar System thereby explaining some anomalous orbits of trans-Neptunian objects \cite{Scholtz:2019csj} would certainly open the possibility to tighten the lower bound of $\Lamstar$ by several orders of magnitude.\footnote{Proposals to identify the conjectural primordial black hole that might be playing the role of Planet 9 in the coming decades have been proposed in the last three months \cite{Witten:2020ifl,lawrence2020bruteforce,Siraj:2020upy,Arbey:2020urq}.} On the purely theoretical side, possible issues with perturbative unitarity, causality or the existence of a well-behaved graviton scattering S-matrix are headache pitfalls \cite{Camanho:2014apa,Chowdhury:2019kaq,deRham:2020zyh}. Further research is needed and is certainly underway.

\section*{Acknowledgements}

We would like to thank Jos\'e Juan Blanco Pillado, Pepe Barb\'on and Enrico Pajer for discussions.
The work of JDE, DVR and AVL is supported by MINECO FPA2017- 84436-P, Xunta de Galicia ED431C 2017/07, Xunta de Galicia (Centro singular de investigaci\'on de Galicia accreditation 2019-2022) and the European Union (European Regional Development Fund -- ERDF), ``Mar\'\i a de Maeztu'' Units of Excellence MDM-2016-0692, and the Spanish Research State Agency.
DVR is supported by Xunta de Galicia under the grant ED481A-2019/115.
AVL is supported by the Spanish MECD fellowship FPU16/06675.
This work was supported in part by the Natural Sciences and Engineering Research Council of Canada.
DVR is pleased to thank the University of Waterloo where part of this work was done, for their warm hospitality.

\end{document}